\input{aipcheck}

\documentclass[
    ,final            
  ]
  {aipproc}

\layoutstyle{6x9}

\begin{document}

\title{Pop III Stellar Masses and IMF}


\author{Michael L. Norman}{
  address={Center for Astrophysics and Space Sciences and San Diego Supercomputer Center, 9500 Gilman Dr., University of California San Diego, 92093}
}

\begin{abstract}
We provide a status report on our current understanding of the mass scales for Pop III.1 and Pop III.2 stars. Since the last review \cite{Norman08}, substantial progress has been made both numerically and analytically on the late stages of protostellar cloud core collapse, protostar formation and accretion, and stellar evolution taking into account cloud core properties and radiative feedback effects. Based on this, there are growing indications that primordial stars forming from purely cosmological initial conditions (Pop III.1) were substantially more massive than stars forming in preionized gas (Pop III.2) where HD cooling is important. Different stellar endpoints are predicted for these two types of Pop III stars with different chemical enrichment signatures: the former die as pair instability supernovae or intermediate mass black holes, whereas the latter die as iron core-collapse supernovae, leaving behind neutron star and stellar black hole remnants. We review recent simulations which show evidence for binary fragmentation at high densities, and comment on the significance of these results. We then summarize an attempt to directly calculate the Pop III.1 IMF taking into account the latest numerical and analytical models. We conclude with suggestions for the kind of simulations needed next to continue improving our understanding of Pop III star formation, which is a necessary input to understanding high redshift galaxy formation. 
\end{abstract}

\maketitle


\section{Introduction}
The standard model of Pop III star formation holds that they are born massive ($\sim 100 M_{\odot}$) and form as isolated, single stars in the centers of dark matter minihalos of mass $\sim 10^6 M_{\odot}$  at redshifts z=30--15 (\cite{Norman08} and references therein.) The mass scale for Pop III stars is fundamentally linked to the cooling properties of primordial gas that virializes to $T \sim 1000$K in such halos. Trace amounts of $H_2$ form via the $H^{-}$  channel which cools the densest gas to $\sim 200$K---a temperature set by the excitation temperature of the lowest rovibrational transition. At densities of $\sim 10^4 cm^{-3}$  the $H_2$ level populations of assume Boltzmann statistics and cooling becomes independent of density. The Bonner-Ebert mass for gas of this density and temperature is about $500 M_{\odot}$. Once $H_2$ cooling has accumulated this much mass at these conditions in the central part of the halo, gravitational collapse ensues. However, due to the low fractional abundance of $H_2$ molecules ($\sim 10^{-4}$), collapse is a quasistatic cooling flow until densities reach about $10^8 cm^{-3}$. At this point, gas becomes fully molecular via the 3-body formation process and the collapse becomes dynamic. What happens next is fundamental to determining the Pop III mass scale and IMF. Does the gas fragment into smaller objects or accrete onto a single massive protostar? What are their characteristic masses and mass distributions?

In this contribution I will review work carried out since my 2008 review \cite{Norman08} that examines the high density evolution of the primordial protostellar cloud core, the formation and growth of the protostar, and stellar evolution through to its endpoint. Recent numerical simulations suggest that massive single Pop III stars and also binary stars are plausible outcomes. Scenarios are also discussed which, depending on additional physics assumptions, may produce more than two fragments in a bound system which opens the possibility of dynamic ejection of a low mass primordial star that never accretes more mass. The role of radiative feedback on setting the final stellar mass is briefly reviewed, but see also Tan et al. (these proceedings). 

\section{Formation of Gravitationally Collapsing Cores}
We adhere to the nomenclature proposed by McKee at First Stars III \cite{OSheaFS3}. Pop III.1 stars are those that form from pristine initial conditions set by cosmology and primordial gas physics. Pop III.2 stars form from zero metallicity gas that has been pre-processed by radiative or mechanical feedback effects, but not by chemical enrichment. Examples of the latter would be stars forming in relic HII regions \cite{OShea05}, in the presence of a strongly photodissociating UV backround \cite{Machacek01,WA07,ON08},
or in shocked primordial gas shells \cite{Mackey03}.
Technically, stars forming from gas of such low metallicity that their formation dynamics are indistinguishable from Pop III are also included in this class.

\subsection{Pop III.1}

A growing list of numerical simulations starting from cosmological initial confirms the basic picture outlined in the Introduction. As reviewed in \cite{Norman08}, both AMR and SPH simulations give results that agree well with one another for the same cosmology, input physics, and numerical resolution up to central densities of $10^{12} cm^{-3}$ where they can be compared. Evolution to higher central densities is still being explored and is discussed below.

The formation timescale and dynamics for cores forming in the presence of a photodissociating UVB is not significantly different from the zero UVB case for $J_{LW} <10^{-23} erg/cm^2/s/Hz/ster \equiv 10^{-2} J_{21}$ \cite{ON08}, and therefore operationally we can include them in Pop III.1. The evolution of the Lyman-Werner (LW) background due to Pop III.1 sources predict that this level is reached at $z \sim 30$ \cite{Y03, WA05}, and thus we can take this as the nominal transition from Pop III.1 to Pop III.2. Note that the transition redshift is sensitive to $\sigma_8$ and the mass of the first stars (through their luminosities). Yoshida et al. \cite{Y03} and Wise \& Abel \cite{WA05} both assumed WMAP1 parameters ($\sigma_8 = 0.9$), which would overestimate the transition redshift based on the WMAP7 value of $\sigma_8 = 0.82$.  It is also important to note that proximity effects (i.e. the suppression of $H_2$ formation and cooling in clustered minihalos) are ignored in these estimates. The highly clustered nature of high redshift structure formation potentially makes this an important effect which would increase the transition redshift. A lower limit to the transition redshift comes from the buildup of the LW background due to normal stars (Pop II) forming in high redshift dwarf galaxies. Ahn et al. \cite{Ahn09} simulated this process in a 50 Mpc box by considering radiation escaping from galaxies with dynamical mass $> 10^8 M_{\odot}$ (atomic line coolers). This simulation explicitly ignores the contribution of minihalos to the LW background. They found that the volume averaged $J_{21}$ reaches $10^{-2}$ at $z \approx 20$, which we can take as an approximate lower limit for Pop III.1 formation given the definition above.

In the absence of a LW background, O'Shea and Norman \cite{ON07} explored the effect of formation redshift on the properties of the collapsing cloud core by simulating 12 cosmological realizations using the AMR code Enzo. They found that halos virializing at higher redshifts ($z>30$) systematically produce lower mass cores than $z \sim 20$ halos because the higher virial temperature for a given halo mass translates into more $H_2$ production and hence more cooling. More cooling leads to lower temperatures and accretion rates in the collapsing core, suggesting that Pop III.1 stars forming at high redshift may be less massive than those forming later. Tan et al. (these proceedings) has investigated the effect this might have on final stellar masses by applying the accretion shutoff model of McKee \& Tan \cite{MT08} (hereafter MT08). They confirm the existence of the effect but find that is it weak and dominated by intrinsic scatter. Implications of this for the Pop III.1 IMF are discussed below. 

\subsection{Pop III.2}

Primordial stars forming from initial conditions modified by radiative and/or mechanical feedback are called Pop III.2 stars. Two cases are particularly important: 1) Pop III stars forming in the presence of a strong LW background $J_{21} > 10^{-2} (z<30)$ from gas that has not been preionized or otherwise disturbed, and 2) Pop III stars forming in pre-ionized gas that has subsequently recombined and formed $H_2$ and HD molecules, such as in a relic HII region. 

In the first case, we imagine a minihalo of characteristic mass $10^6 M_{\odot}$ forming significantly later and far away from than those forming from the rare ($\geq 3\sigma$) peaks which formed Pop III.1 stars. Yoshida et al. \cite{Y03} showed that $H_2$ production in minihalos is suppressed for $J_{21} > 10^{-2}$. Using high resolution AMR simulations O'Shea \& Norman \cite{ON08} (hereafter ON08) showed that collapsing core formation is not eliminated for $J_{21} > 10^{-2}$, but only substantially delayed. They showed that even for values as large as $J_{21}=1$ core collapse eventually happens because although the abundance of $H_2$ is very low at these background levels, the cooling rate per molecule is a strongly increasing function of temperature. At some point the halo becomes massive enough through accretion and mergers that its virial temperature becomes high enough so that the higher cooling rate per particle multiplied by a long evolutionary time ($\sim 200$ Myr) removes enough thermal energy to trigger gravitational collapse. Such ``Lyman-Werner delayed halos" accumulate more mass in the core and of higher temperature (see O'Shea \& Whalen, these proceedings) than Pop III.1 star-forming halos. Once collapse begins in a delayed halo, the accretion rates are higher as well, suggesting that either the Pop III.2 star will be more massive than Pop III.1 stars if only a single star is formed, or that several Pop III.2 stars will form. Calculations of this sort have not been carried to a high enough density to know which of these two possibilities is more likely.
  
The second case---primordial stars forming in recombining gas---is expected to be common, and possibly even the dominant mode of Pop III star formation. The reasons for this are several. First, due to the clustering in CDM models, the second halo to form a Pop III star is very likely to be near the first one to do so in a given volume of the universe. This is borne out by simulations \cite{OShea05,Mesinger06,Y07}.
Second, Pop III stellar lifetimes are only about 1\% of the Hubble time during the Pop III epoch, and recombination times are also short compared to the Hubble time for $z > 10$, meaning HII regions come and go quickly in a given region of space, even as the ionized volume fraction slowly grows. Third, Pop III star formation rate density, and the ionized volume fraction secularly increases toward lower redshifts, meaning that the likelihood a Pop III star forms in this way also increases toward lower redshifts.
 
Primoridal stars forming in relic HII regions form from initial conditions that are thermally, chemically, and structurally altered relative to the Pop III.1 case.  The most careful simulations of this scenario were carried out by Mesinger et al. \cite{Mesinger06} and Yoshida et al. \cite{Y07}. They found that the high electron abundance in the recombining gas catalyzed rapid formation of $H_2$ molecules which roughly compensated the Jeans smoothing of the gas during the ionized phase.  Ignoring the effects of HD formation and cooling, they found that core properties and accretion rates were very similar to the Pop III.1 case. However, when HD cooling was included, they found lower core temperatures (100K vs. 200K) which lowered the core accretion rates substantially. Simple accretion time arguments suggested typical final stellar masses of $40-60 M_{\odot}$, rather than $100-300 M_{\odot}$ characterizing Pop III.1 stars. 

\section{From Collapsing Cores to Protostars}
Additional physical processes must be included to evolve the collapsing cloud core to stellar density. These include: 1) heat of formation of $H_2$ and change of the adiabatic index above $n \sim 10^{8} cm^{-3}$; 2) radiative transfer of the $H_2$ cooling lines above $n \sim 10^{12} cm^{-3}$; 3) radiative cooling due to collisionally induced emission (CIE) above $n \sim 10^{14} cm^{-3}$ including continuum opacity effects; and 4) cooling due to $H_2$ dissociation above $n \sim 10^{15} cm^{-3}$. Omukai \& Nishi (1998)\cite{ON98} were the first to include all of these processes in 1D spherically symmetric collapses. A decade later, Yoshida and colleagues achieved it in 3D \cite{Y08} starting from proper cosmological initiation conditions--truly a milestone in the subject. Radiative transfer effects were modeled using an escape probability formalism calibrated against full radiative transfer calculations in 1D.

Fig. 1 shows structures on four different lengthscales from the final snapshot of their simulation which spans a range of scales of 13 orders of magnitude! The newly elucidated structures in this simulation are the centrifugally supported, Toomre unstable disk in Fig. 1c, and the hydrostatic central protostar in Fig. 1d. The latter is formed with a mass of $10^{-2} M_{\odot}$  and is accreting at a rate between 0.01 and 0.1 $M_{\odot}/yr$, implying that it will grow to a mass of tens of solar masses in just $10^4$ yr. The simulation, which resolves the Jeans mass everywhere, shows no evidence for fragmentation.

\begin{figure}
  \includegraphics[width=.8\textwidth]{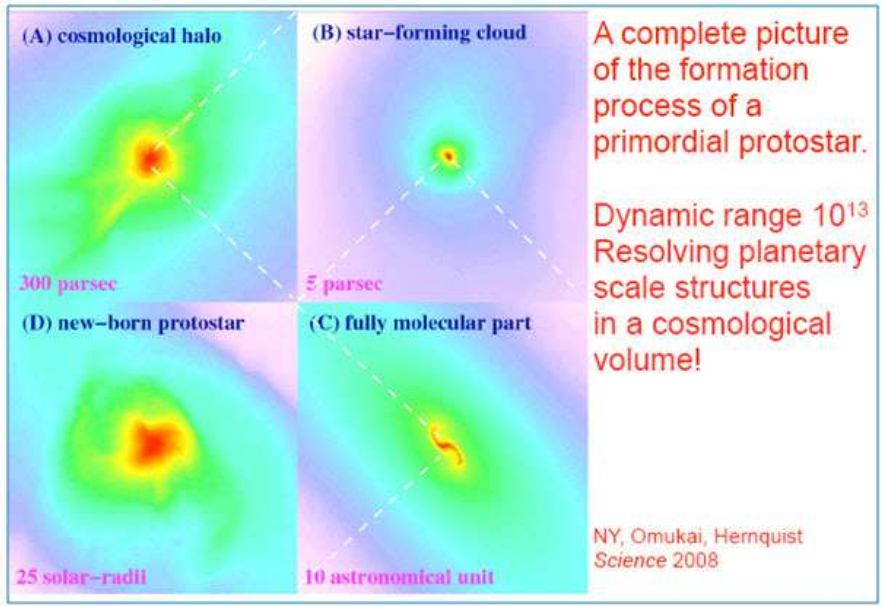}
  \caption{A zoom-in showing structures on diverse length scales from the first 3D
hydrodynamic cosmological simulation to form a hydrostatic, accreting primordial
protostar. From \cite{Y08}.}
\end{figure}

Time-explicit numerical simulations such as this cannot follow the accretion phase because of the very small timestep imposed by the Courant condition. However, the accretion rate history onto the central protostar can be estimated from a snapshot of the mass density and velocity distributions in the envelope, and then fed into a 1D protostellar evolution code. Ignoring late stage fragmentation, Yoshida et al. \cite{Y07} estimate a Pop III.1(III.2) enters the ZAMS with a mass of $\sim 100(40) M_{\odot}$, respectively. 

It is important to note that Pop III stars continue to accrete after the onset of hydrogen burning. Their final mass will be set either by the mass available to accrete, or by radiative feedback effects \cite{MT08}. Interestingly, in the case of a Pop III.2 star forming in a relic HII region, \cite{Y07} found that the cloud core was only about $ 40 M_{\odot}$, suggesting that this number may be close to the final stellar mass. Pop III.1 stars, and Pop III.2 stars forming in Lyman-Werner delayed halos are surrounded by much more massive cloud cores and may accrete until radiation terminates the accretion. This is discussed further below.

\section{Fragmentation}
\subsection{Core fragmentation due to turbulence}

The collapsing cores of primordial minihalos are characterized by a state of subsonic turbulence, and therefore there is a ready source of perturbations to trigger fragmentation if the gas is thermally or chemo-thermally unstable \cite{Silk83}.  Fragmentation was not observed in the cosmological simulations of Abel et al. \cite{ABN00,ABN02} and Yoshida et al. \cite{Y06,Y08}, although it was observed in the idealized collapse simulations of Bromm et al. \cite{BCL99} and Clark et al. \cite{Clark10}. Clearly, initial conditions matter. We still have rather small samples of self-consistent, high dynamic range simulations of Pop III star formation which inform our standard model. O'Shea \& Norman \cite{ON07} simulated four cosmic realizations in each of three different box sizes and found no fragmentation up to central densities of $10^{12} cm^{-3}$. However, Turk et al. \cite{Turk09} found one case out of 5 higher density evolutions in which the core fragments into two subcores, each of which were in a state of freefall collapse. With a separation of nearly 1000 AU, Turk et al. suggest the final outcome is a long period Pop III binary. Turk (these proceedings) reports a 20\% binary incidence rate in a larger sample of 25 simulations. 

\subsection{Late fragmentation due to disk instabilities}
The formation of a rotationally supported, fully molecular disk as in Fig. 1c creates an environment for fragmentation. The disk will fragment if it becomes Toomre unstable.  Toomre instability depends on the gas surface density $\kappa$ through the Q parameter: $Q=c_s \kappa / \pi G \Sigma$ where $\kappa$ is the epicyclic frequency. If material is added to the disk faster than it can be deposited onto the central object, the surface density will grow to the point where the disk becomes gravitationally unstable. The strong $m=2$ mode seen in Fig. 1c is the first manifestation of global Toomre instability. If Q drops substantially below unity, the disk becomes locally gravitationally unstable and can fragment into multiple objects.

A technical impediment to exploring possible disk fragmentation in a numerical simulation like that shown in Fig. 1 is the telescoping down of the timestep as the central object becomes denser and hotter. One approach to avoiding the ``Courant catastrophe" is to replace the central object with a sink particle, thus regularizing the simulation. Sink particles were introduced by Bate et al. \cite{Bate95} in the context of Galactic star formation simulations, and first employed for primordial star formation by \cite{Bromm04}. At an operational level, all gas above a given density is converted into a sink particle thus placing a floor on the timestep. The sink gains mass from its environment as gas particles cross its gravitational radius. 

Stacy et al. \cite{Stacy10} employed sink particles to study the evolution of the accretion disk surrounding a Pop III.1 protostar in a standard cosmological setup. They chose as their threshold density for creating a sink $10^{12} cm^{-3}$ by which time the gas is fully molecular but before the additional physical processes mentioned above become important. They integrated for 5000 yr after sink creation, during which time the sink grew in mass from 0.7 to 43 $M_{\odot}$. They terminated the calculation before radiative feedback effects become important. They found that a centrifugally supported disk forms around the protostar which grows to a size of $\sim 2000$ AU and mass $\sim 40 M_{\odot}$ in 5000 yr. The disk fragments into 5 objects, two of which are massive and long-lived (the central star and one at 700 AU). 

One needs to be careful accepting this result at face value. The first concern is the sensitivity of the results to the threshold density for making a sink. A higher/lower density will likely result in a smaller/larger disk (this effect has been seen in AMR simulations, where an artificial pressure is used to regularize the simulation.) Assuming the disks fragment in a similar way, the separation of the multiple protostars would be sensitive to the threshold density. This is borne out by the higher resolution simulations presented by Clark (these proceedings) which used a threshold density of $10^{15} cm^{-3}$. The second concern is the correctness of the dynamical evolution of the sink particles once formed. While a sink particle treatment for the central protostar seems ``safe" because it sits in the center of the potential well and accretes rapidly, condensations forming in the disk are subject to very different stresses (gravitational torques, hydrodynamic shear). Converting extended condensations into sinks gives them an indestructability that may not be realistic. Moreover, whether the condensations migrate inward and accrete onto the central star, or migrate outward to become a stable binary/multiple system depends on modeling the torques and hydrodynamic drag accurately. Despite these caveats, however, these simulations do lend strong support to the notion that at least some and perhaps all Pop III stars form as binaries. 

\begin{figure}
  \includegraphics[width=.5\textwidth]{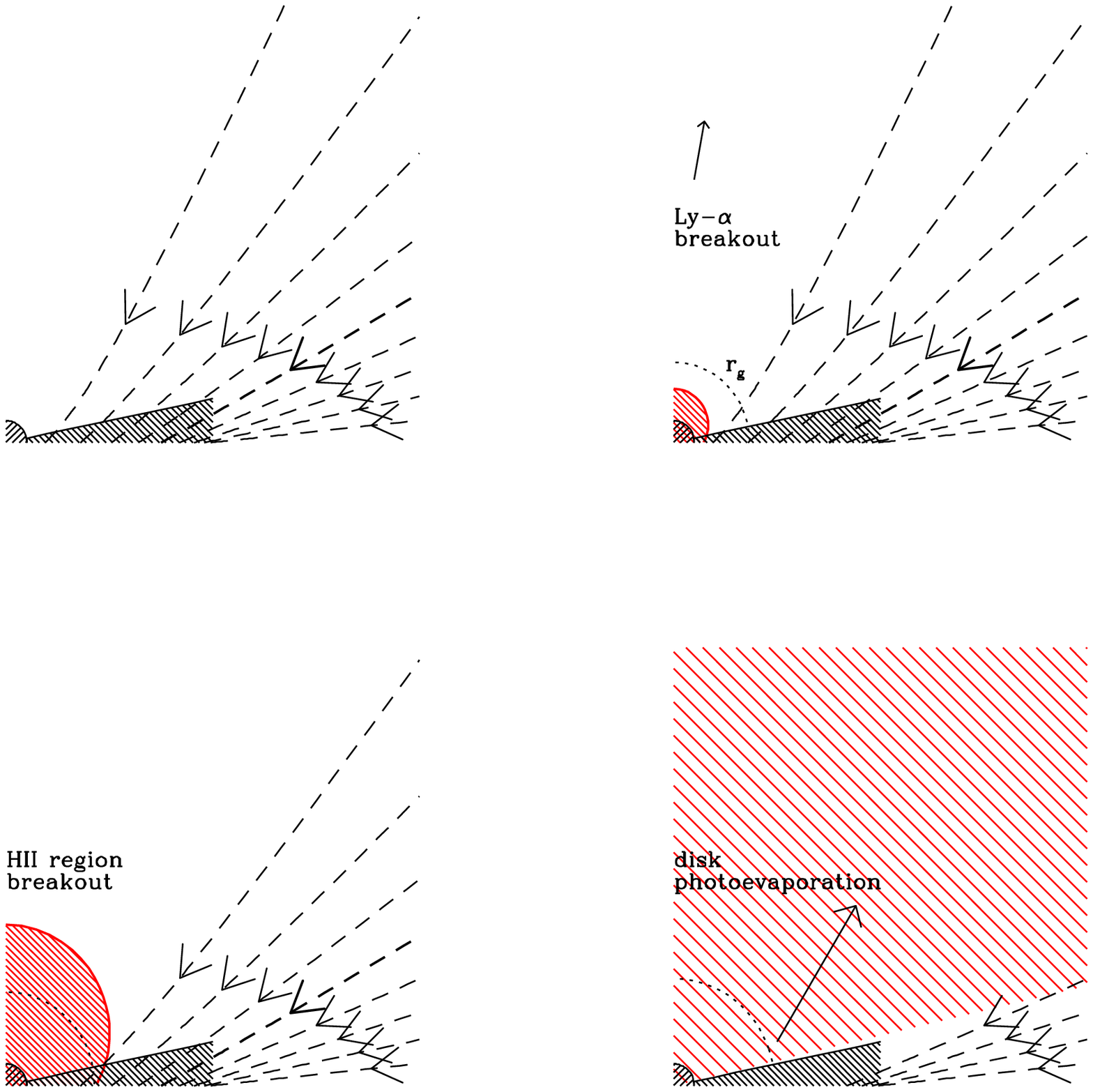}
  \includegraphics[width=.5\textwidth]{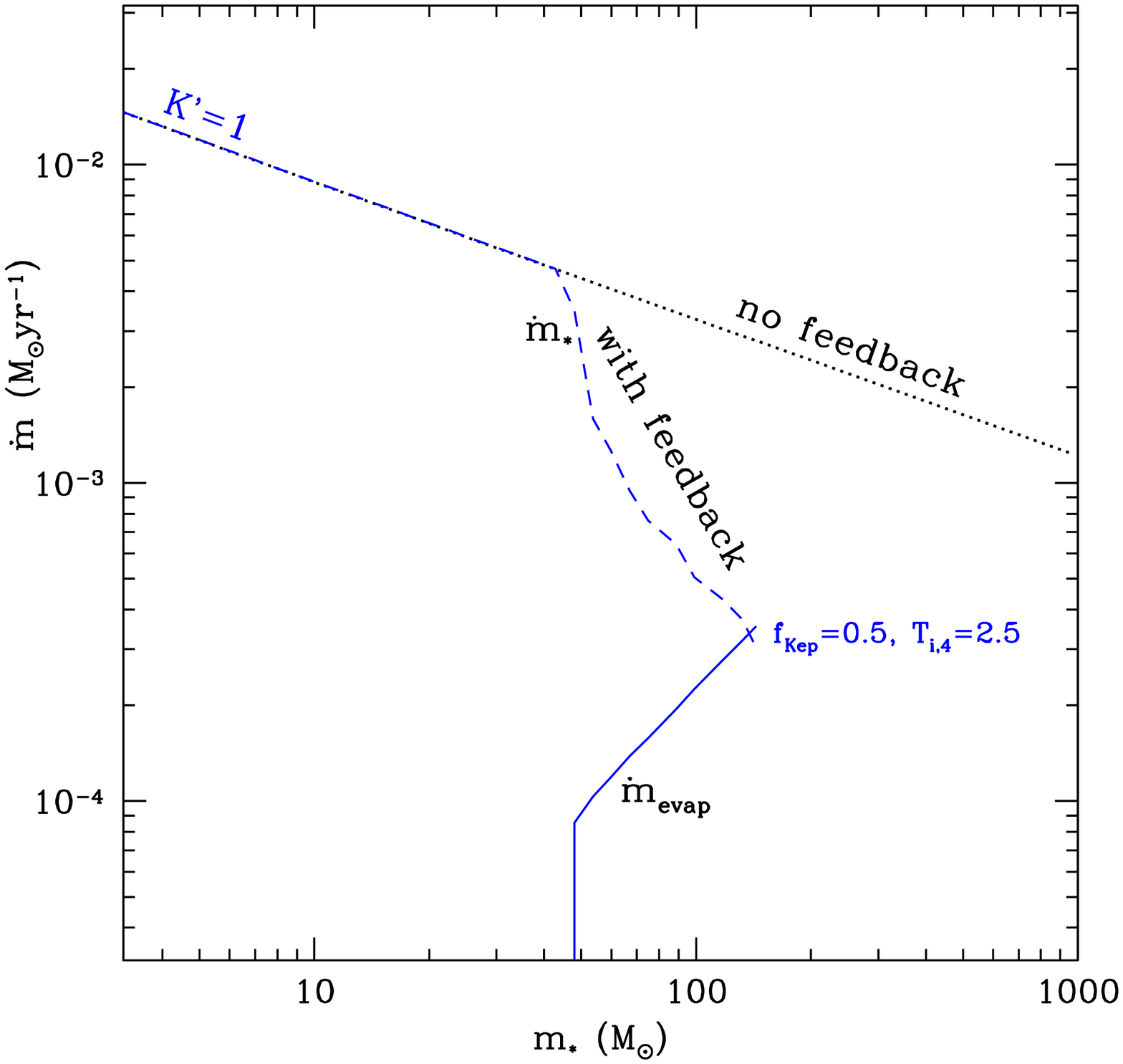}
  \caption{{\em Left:} Schematic of the evolutionary phases of an accreting Pop III
protostar. {\em Right:} Protostellar accretion rate versus accreted mass including
radiative feedback. When the disk photoevaporation rate equals the accretion rate, 
accretion stops. From \cite{MT08}.}
\end{figure}

\section{From Protostars to Stars to Stellar Endpoints}
\subsection{Halting accretion: radiative feedback}
Because Pop III stars are predicted to be ultraluminous \cite{Schaerer02}, it is important to consider the effects of radiation on the accreting envelope.  McKee \& Tan \cite{MT08} studied this within the context of analytic models of Pop III star formation. The model assumes that the Pop III protostar accretes from a disk whose properties are estimated from the conditions found in the collapsing cloud cores from the cosmological simulations of Abel et al. \cite{ABN02} (hereafter ABN02). They considered various forms of radiative feedback, including $H_2$ photodissociation, Lyman $\alpha$ and Lyman continuum pressure, HII region breakout, and accretion disk photoevaporation. They found that the first three of these effects are not significant at any time, and the last two only become important at later times as the protostar becomes more massive and luminous. 

Fig. 2 illustrates the basic mechanism that shuts off disk accretion. As the accretion rate from the envelope diminishes, and the star's UV luminosity rises, an HII region breaks out above and below the disk. The expanding HII region blocks accretion above and below the disk, reducing the accretion rate below the no feedback value, and bathes the disk in ionizing radiation from recombinations within the nebula. The disk begins to photoevaporate. When the disk photoevaporation rate equals the accretion rate, inflow is halted and the star attains its final mass. The final mass depends on the entropy of the gas accreting from large radius, as well as its specific angular momentum. Fig. 2b shows the fiducial case from MT08 taking values from the Pop III.1 simulation of ABN02. They predict a final mass of $\sim 140 M_{\odot}$.

\subsection{Pop III stellar evolution}
Ohkuba et al. \cite{Ohkubo09} used accretion histories from the cosmological simulations of \cite{Y07,Y08} to carry out Pop III.1 and Pop III.2 stellar evolution calculations through to their end points. To consider some variation in the parent cloud, they parameterize angular momentum and the radiative feedback effects of MT08. Ignoring radiative feedback, they find Pop III.1 stars die as very massive stars (VMS) in the range $300 - 1000 M_{\odot}$, depending on the angular momentum of the cloud. These stars produce intermediate mass black hole (IMBH) remnants and no chemical enrichment. Note this mass range does not include the range $140 - 260 M_{\odot}$ predicted to produce pair instability supernovae (PISN)\cite{Heger02}. However, when radiative feedback is included, they find that Pop III.1 stars die with masses in the range $60 - 320 M_{\odot}$, depending on angular momentum of the cloud.  This mass range includes Type II supernovae, PISN, and IMBHs as possible outcomes. Clearly, radiative feedback effects are fundamental in determining the Pop III IMF and chemical signatures. Finally, they find that Pop III.2 stars die as core-collapse supernovae in the mass range  $40 - 60 M_{\odot}$, consistent with earlier estimates \citep{Mesinger06,Y07}. These stars produce stellar black holes and some chemical enrichment.  

\begin{figure}
  \includegraphics[width=.5\textwidth]{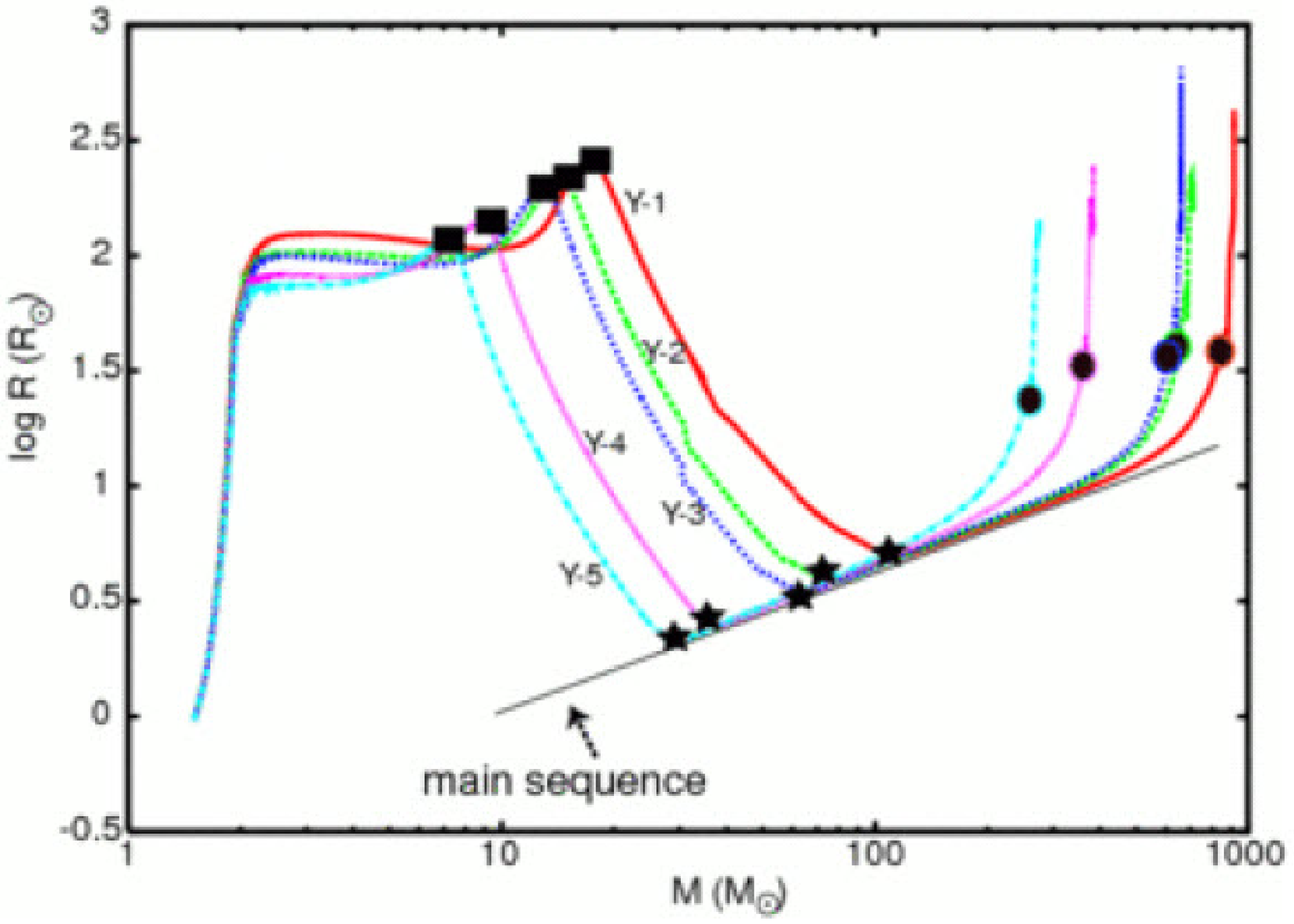}
  \includegraphics[width=.5\textwidth]{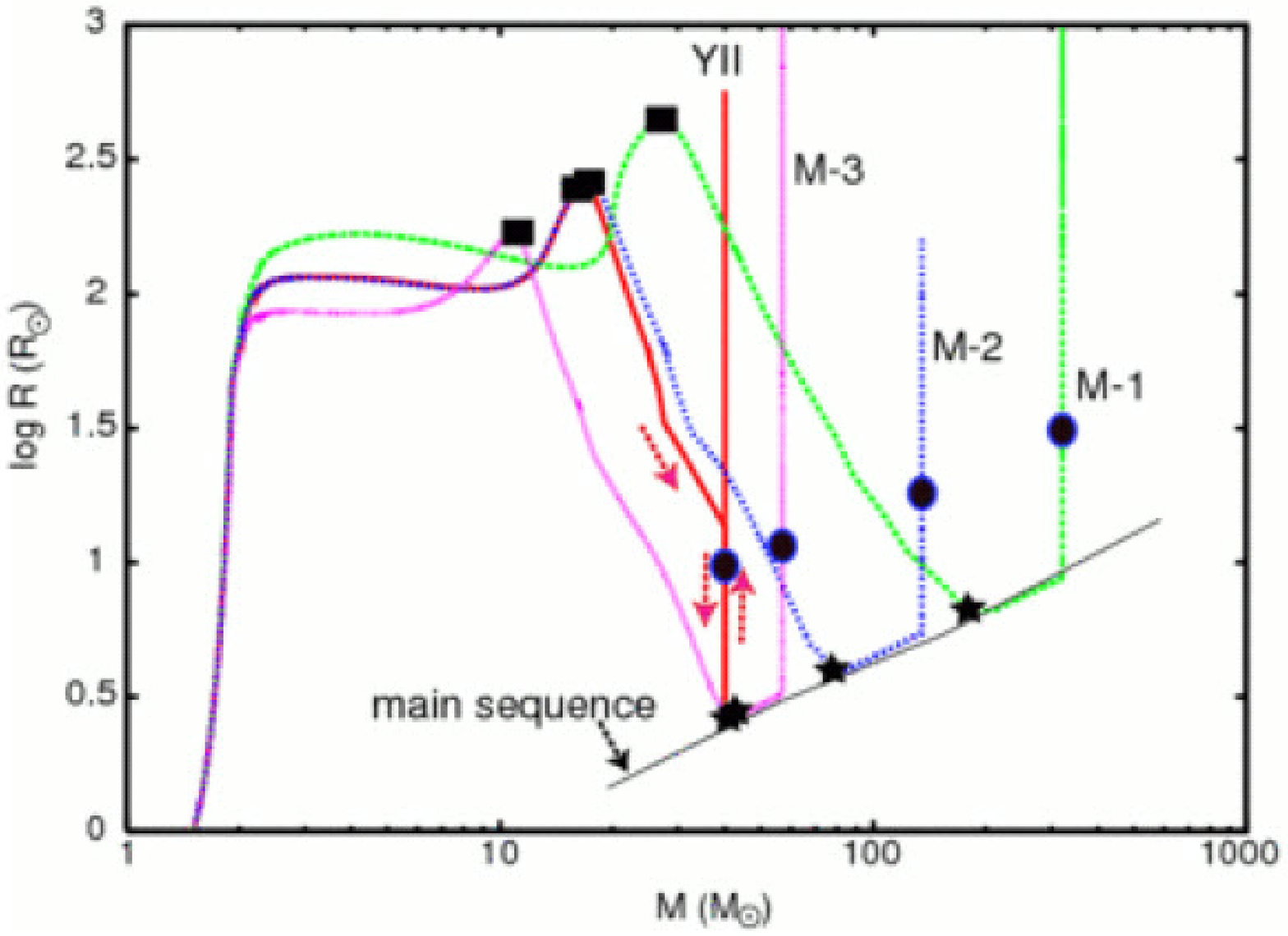}
  \caption{Evolution of Pop III protostars and stars to their endpoints.
{\em Left:} without radiative feedback; {\em Right:} with radiative feedback.
Cases Y-1 to Y-5 are Pop III.1 protostars with accretion rates reduced by varying amounts 
due to rotation. Case YII is a Pop III.2 protostar. Cases M-1 to M-3 are Pop III.1
stars varying the ionized gas temperature and accretion rate. From \cite{Ohkubo09}.}
\end{figure}

Given the findings described above, what can we say about the Pop III IMF? Ignoring binarity for the moment, it is now quite clear that Pop III.1 stars are considerably more massive than Pop III.2 stars forming in relic HII regions. Table 1 in MT08 quote values of the photoevaporation limited mass in the range of $\sim 60 - 320 M_{\odot}$ varying parameters over a reasonable range, with a fiducial mass of $140 M_{\odot}$. Ohkubo et al. \cite{Ohkubo09} find similar values for the stellar endpoint masses, although this is not surprising since they used fits to MT08 rates as input to their calculations.   Tan et al. (these proceedings) used the ensemble of 12 simulations of \cite{ON07} to estimate the Pop III.1 IMF. They confirmed that stars forming at lower redshift are more massive, but caution that the trend is weak and there is a large scatter. The key factor setting the final mass is the entropy parameter for the gas in the Jeans unstable core.  High masses are obtained in higher temperature cores. They predict final masses in the range $80 - 600 M_{\odot}$, with a mean(median) value of $250(215) M_{\odot}$, respectively. Based on this small sample, they compute a skewed IMF that has a peak at $\sim M_{\odot}$, falling off very rapidly to lower masses, and more gradually to higher masses. 

Although the sample of cosmological simulations of Pop III.2 stars forming in relic HII regions is even smaller \citep{OShea05,Mesinger06,Y07},
there is a consensus that a lower value of $40 - 60 M_{\odot}$ is more appropriate. This mass reflects the smaller mass of the cloud core to accrete from, and the lower accretion rates that results from HD cooling the gas to $\leq 100$K. At these mass scales, radiation feedback is less important in setting the mass scale. 

Pop III.2 stars forming in Lyman-Werner delayed halos should fall in the mass range of the Pop III.1 stars mentioned above, and possibly extend it to higher masses \cite{ON08}. The reason is that delay translates into higher virial masses and temperatures for the minihalo before gravitational instability sets in. Once it sets in, accretion rates are higher. ON08 observe a discontinuous jump in the virial mass/temperature of the minihalo at the onset of gravitational instability between $J_{LW}=10^{-1.5} J_{21}$ and $10^{-1} J_{21}$  (see Fig. 1 of O'Shea, these proceedings, or Fig. 3 of ON08.) These values are ($2.5 \times 10^6 M_{\odot}, 4000$K) and ($10^7 M_{\odot}, 9000$K), respectively. The mass of the collapsing cores when the central density reaches $10^{10} cm^{-3}$ are $3\times 10^3 M_{\odot}$ and $6\times 10^3 M_{\odot}$, respectively. Provided such objects can form in the universe, we might expect a discontinuous change in the characteristic stellar masses that are formed at these values of $J_{LW}$. Once collapse begins, the $H_2$ fraction at the onset of collapse becomes unimportant, only the mass and entropy of the collapsing core. Although no fragmentation was seen in the simulations to a central density of $10^{10} cm^{-3}$, we have seen that fragmentation can occur at higher densities. With the larger supply of infalling material, it is possible that the outcome is a small cluster of $\sim 100 M_{\odot}$ stars. 

The possibility of binary formation does not change the basic picture outlined above, but merely divides the mass scales by two. If final masses are determined by the amount of gas available to accrete onto the binary protostars, as in relic HII regions, then masses of $20 - 30 M_{\odot}$ would be predicted. However, Pop III.1 stars and Pop III.2 stars forming in Lyman-Werner delayed halos have a larger reservoir of gas to accrete from. Final masses may be set by a complex interplay between radiative feedback and high density fragmentation effects. More gas may simply mean more stars of the characteristic mass $\sim 100 M_{\odot}$ set by photoevaporation limited disk accretion. 

\section{What Simulations are Needed Now?}

We need larger samples of cores from cosmological initial conditions reaching central densities of at least $10^{12} cm^{-3}$ in various Lyman-Werner fluxes to explore the turbulent state of the gas and core multiplicities, as well as to quantify the entropy and rotational parameters entering in the MT08 model. From this sample an improved estimate of the Pop III.1 IMF can be obtained using the procedures of Tan et al. (these proceedings). 

We need a statistical sample of cores from cosmological initial conditions forming in relic HII regions, taking the detailed minihalo photoevaporation dynamics and Lyman-Werner background into account. Idealized studies of minihalo photoevaporation including multifrequency radiative transfer and nonequilibrium primordial gas chemistry by Whalen et al. \cite{Whalen08,Whalen10} show a complicated set of outcomes depending upon the Pop III.1 star's luminosity and spectrum, and the Pop III.2 halo's distance and evolutionary state (O'Shea \& Whalen, these proceedings).  Competing effects (e.g., $H_2$ photodissociation, X-ray preionization and $H_2$  formation, and hydrodynamic cloud crushing) roughly cancel out in many cases, leaving neighboring minihalos is a state of collapse after the I-front has swept over it. Given the expectation that most of Pop III stars from in the vicinity of other Pop III stars and other ionizing sources, this is a rich class of models that need further investigation. 

Sink particles are clearly a valuable tool in investigating the late time, high density evolution of the central region of the cloud. A careful study of disk formation and fragmentation varying resolution and sink recipes would be invaluable to improving our confidence in the results, and elucidating the question of stellar multiplicity. The simulations by Clark et al. (these proceedings) are a step in the right direction.
 
3D radiation hydrodynamical simulations of MT08 model coupled to a radiating protostar evolution model modeled as a sink particle would be an important next step in understanding the details of accretion shutoff. The simulations would be analogous to the massive star formation simulations of Krumholz et al. \cite{Krumholz07}, but include the transport of ionizing radiation. These simulations could profitably be done in idealized setups or in fully cosmological settings. 

Finally, coupled models of the evolution of the Lyman-Werner background and hydrogen reionization in cosmological volumes with adequate resolution to resolve the minihalo population would be numerically challenging but extremely valuable to assess the duration of the Pop III.1 and Pop III.2 epoch, and the impact of primordial stars of all types on the initial conditions for galaxy formation.  


\begin{theacknowledgments}
I thank the conference organizers Volker Bromm, Dan Whalen, and Naoki Yoshida for their invitation to present a review on this topic, and then patiently waiting for my manuscript to be completed. I would like to thank my colleagues Chris McKee, Takuya Ohkubo, Jonathan Tan, and Naoki Yoshida for allowing me to reproduce figures from their publications. I would like to thank my collaborators Tom Abel, Greg Bryan, Brian O'Shea, Britton Smith, Matt Turk, Dan Whalen, and John Wise for keeping it interesting. This work was partially supported by NSF grants AST-0708680, AST-0808184 and NASA grant NNX08AH26G. 
\end{theacknowledgments}



\bibliographystyle{aipproc}   


\end{document}